\documentclass[iicol,pdflatex,sn-mathphys-num]{sn-jnl}% Math and Physical Sciences Numbered Reference Style

\usepackage{graphicx}%
\usepackage{multirow}%
\usepackage{amsmath,amssymb,amsfonts}%
\usepackage{amsthm}%
\usepackage{mathrsfs}%
\usepackage[title]{appendix}%
\usepackage[dvipsnames]{xcolor}%
\usepackage{textcomp}%
\usepackage{manyfoot}%
\usepackage{booktabs}%
\usepackage{algorithm}%
\usepackage{algorithmicx}%
\usepackage{algpseudocode}%
\usepackage{listings}%
\usepackage{gensymb}

\theoremstyle{thmstyleone}%
\theoremstyle{thmstyletwo}%

\theoremstyle{thmstylethree}%f

\begin{document}

\title[Article Title]{U-Net based particle localization in granular experiments: Accuracy limits and optimization}

\author*[1,2]{\fnm{Fahad} \sur{Puthalath}}\email{fahad.puthalath@dlr.de}
\author*[3]{\fnm{Matthias} \sur{Schr\"oter}}\email{matthias.schroeter@phys.uni-goettingen.de}
\author[1,4]{\fnm{Nicoletta} \sur{Sanvitale}}%\email{nicoletta.Sanvitale@dlr.de}
\author[1,2]{\fnm{Matthias} \sur{Sperl}}%\email{matthias.sperl@dlr.de}
\author*[1]{\fnm{Peidong} \sur{Yu}}\email{peidong.yu@dlr.de}

\affil*[1]{\orgdiv{Institut f\"ur Frontier Materials auf der Erde und im Weltraum}, \orgname{Deutsches Zentrum f\"ur Luft- und Raumfahrt (DLR)}, \orgaddress{\city{K\"oln}, \postcode{51170}, \country{Germany}}}

\affil[2]{\orgdiv{Institut f\"ur Theoretische Physik}, \orgname{Universität zu K\"{o}ln}, \orgaddress{\street{Z\"ulpicher Strasse 77}, \city{K\"oln}, \postcode{50937}, \country{Germany}}}

\affil[3]{\orgdiv{Institute for the Dynamics of Complex Systems, University of G\"ottingen}, \orgaddress{\street{Friedrich-Hund-Platz 1}, \city{G\"ottingen}, \postcode{37077}, \country{Germany}}}

\affil[4]{Institute of Radio Astronomy, INAF, Italy}
%\affil[4]{\orgdiv{Istituto di Radio Astronomia, INAF}, \orgaddress{\street{Via P. Gobetti 101}, \city{Bologna}, \postcode{40129}, \country{Italy}}}

\abstract{Identifying the positions of granular particles from experimental images is often complicated by their partial overlap in two dimensional projections. Uneven backgrounds and inhomogeneous illuminations can add to the challenge.
Conventional image-processing methods are often unable to analyze such images. We show that a deep neural network with an U-Net architecture can provide precise particle positions with a high detection rate. For our challenging test image set, the network correctly identifies 97.5\% of the particles while producing only 3.5\% false positives.  The training of the U-Net requires a number of target images where the position of all particles have been identified by humans. Those positions are then indicated in the target images by setting a small number of mask pixels to white in an otherwise black image. We demonstrate that the design of these masks critically determines performance: mask size controls the resolution of overlapping particles, anti-aliased masks enable subpixel accuracy, and systematic human labeling biases set a measurable lower bound on achievable precision.
Our final network achieves an accuracy of the particle coordinate of 3.6\% of the particle diameter.}




\maketitle

\section{Introduction}
\label{sec:intro}

\begin{figure*}[t]
    \centering
    \includegraphics[width=2\columnwidth]{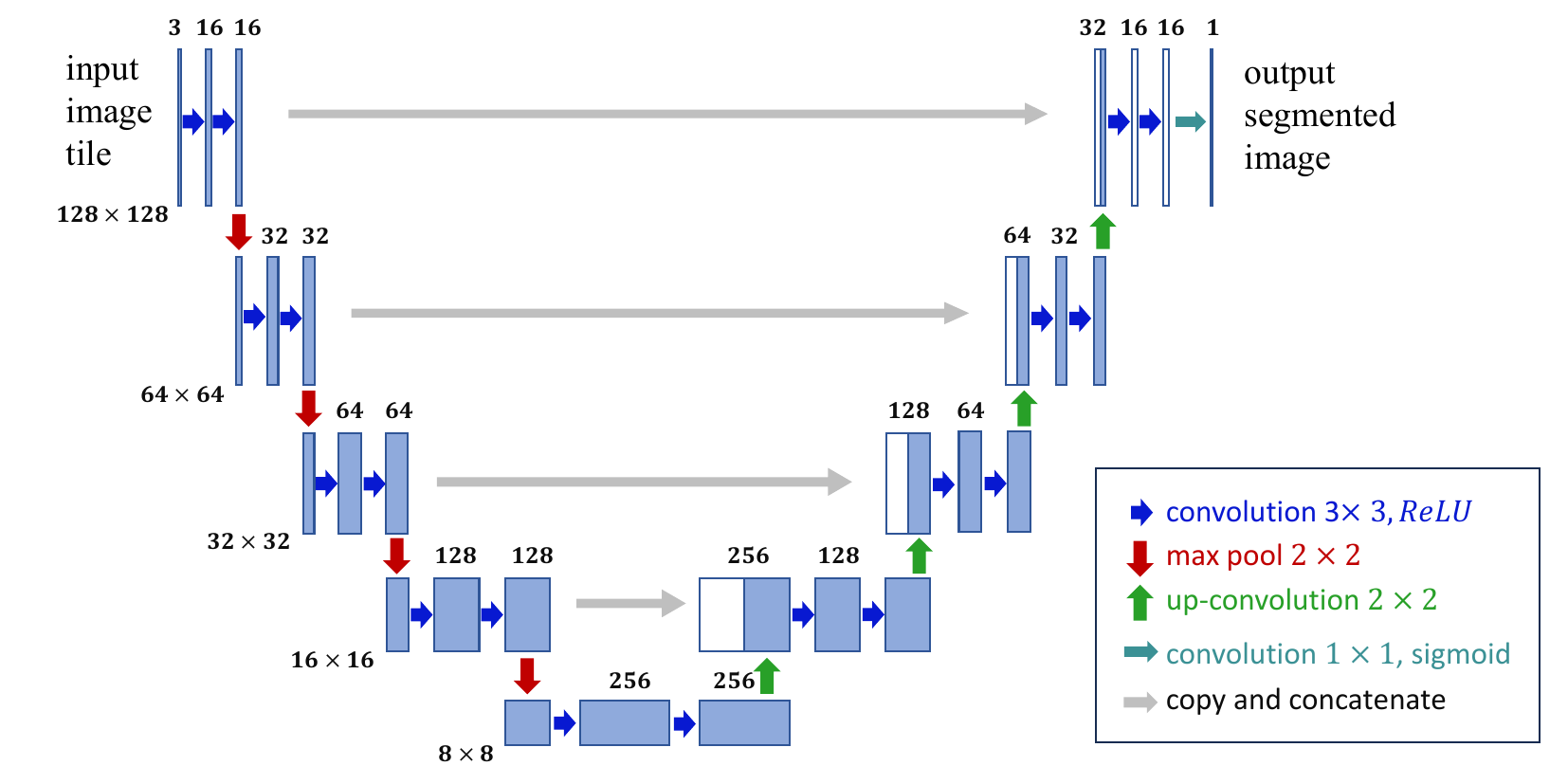}
    \caption{The U-Net architecture adopted in our study. The arrows show the flow of information. The \textit{contraction path} (red arrows) and the \textit{expansion path} (green arrows) form an U-shaped information bottleneck that forces the neural network to learn semantic content.
    Additionally, there are the \textit{copy and concatenation paths} (gray arrows) which preserve the spatial resolution of the original image. The numbers on top of the blocks (and their width) indicate how many feature maps of that size are implemented.}
    \label{fig:Unet}
\end{figure*}

Experimental investigations of granular fluids often require particle tracking for the measurement of various statistical quantities, such as the spatial and velocity distributions, the mean squared displacement, the diffusivity, etc. \citep{Falcon1999, Sack2013, falcon:13, Harth2018, Yu2020, Schneider2021, Pitikaris2022,puzyrev:24, cheng:24}. The first step in particle tracking is the identification of the individual particles and a precise determination of their position within the image.  The task of identifying all image pixels belonging to the particles is called segmentation and it comes in two flavors: semantic segmentation, which means to simply mark all pixels belonging to any of the particles, and instance segmentation, where the pixels belonging to each particle are identified by a different label. The second step in particle tracking is the assignment of the labeled instances in a sequence of images to the same physical particle. This paper does not discuss this second step. But readers interested in innovative assignment methods for sand grains (instead of the spheres discussed here) in dense packings (instead of granular gases) measured in three dimensions using X-ray tomography (instead of two-dimensional camera images) should consult references \cite{cheng:18,cheng:20_b,cheng:21}.

There are two main challenges that the segmentation of granular gas images needs to overcome: first, the three-dimensional nature of the sample results in overlapping particles even in the case of a dilute system.
This is particularly problematic for instance segmentation techniques. Second, in order to enable free particle motion, such granular experiments are often performed in low-gravity conditions such as inside a sounding rocket or a drop tower experiment. The illumination conditions in the confined spaces of these setups are often sub-optimal \citep{Yu2019} which means that the same particle will appear differently in different regions of the sample container. Classical image processing techniques have difficulties to handle this type of image material, c.f.~\cite{Puzyrev2020} or Section \ref{sec:classical}. On the other hand, it has been shown that specialized Convolutional Neural Networks (CNNs) can provide excellent segmentation results on complicated images.  A good example is their great success in the biomedical sciences \citep{ Newby2018, Midtvedt2021, Stringer:21,  Liu2022, weigert2022}.

The use of CNNs in the detection of granular particles under microgravity has been pioneered by Puzyrev \textit{et al}.~\cite{Puzyrev2020,niemann:25} who use Mask R-CNN~\citep{He2017} for tracking rod-shaped granular particles. Mask R-CNN is a two stage algorithm which first identifies a bounding box for each particle and then performs an instance segmentation within that bounding box.  A different approach to segmentation is training a generic CNN on multi-resolution image tiles to determine whether the center pixel of the stack is inside or outside the considered objects \cite{dillavou:24}.

Most of the deep learning segmentation pipelines for granular media are based on the  U-Net architecture  which was initially developed for the segmentation of biomedical images ~\citep{Unet2015}. U-Nets transform input images into output images of the same size that directly represent the segmentation result. Figure \ref{fig:Unet}
shows the characteristic U-shaped architecture that gives the network its name. U-Nets have been used to identify the positions of tracer particles in fluid dynamics experiments \cite{liang:23}, the outlines of stress-birefringent disks \citep{Sanvitale2022}, for the extraction of three-dimensional coordinates from holographic images of particle distributions \cite{shao:20}, and for the segmentation of rock fragments
\cite{liu:20,wang:21,yu:25}.

This paper describes the U-Net architecture we employed to locate spherical metal particles inside a spherical container used in a drop tower experiment. Section 2 introduces our experimental data and expounds the impossibility to analyze them with classical image processing techniques. Section 3 describes the technical details of our machine learning work-flow. Section 4 discusses important aspects and considerations for preparing the target images used for the training, including the influence of the human labelers. Section 5 describes the influence of different choices of hyper-parameters when training the U-Net. We then close with a summary and conclusion.

This paper is accompanied by the source code of our network, its weights, and the data set used for training, validation, and testing\cite{github}.

%===========================================================================================================

\section{Experimental data and classical image analysis}
\label{sec:experiment}

\subsection{Experimental setup}
\label{sec:exp}

\begin{figure}[t]
    \centering
    \includegraphics[width=\columnwidth]{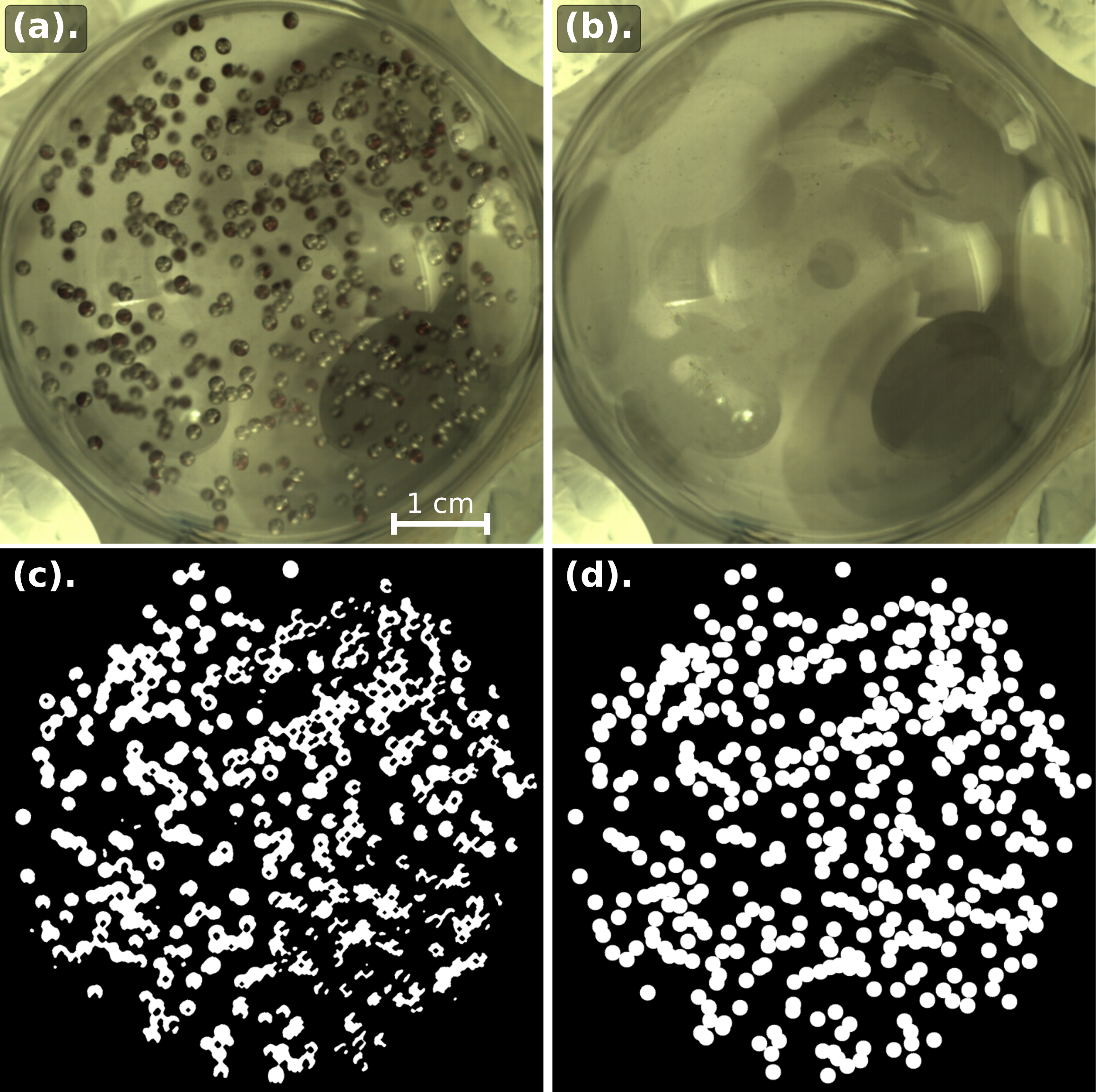}
    \caption{Deep learning identifies the particle positions significantly better than classical image processing methods:
    \textbf{(a)} Raw image from the drop tower experiment.
    \textbf{(b)} Synthetic background image illustrating inhomogeneous illumination and container reflections.
    \textbf{(c)} Semantic segmentation (white pixels indicate particles) via classical image processing.
    \textbf{(d)}  Instance segmentation produced by a trained U-Net.}
    \label{fig:im1_comp}
\end{figure}

The images analyzed in this paper are gathered in an experimental study of granular gases in low gravity. During a typical experiment, approximately 450 spherical magnetic particles of 1.6 mm in diameter are magnetically shaken in a transparent spherical sample cell of diameter 60 mm. The experiment is conducted in a drop tower in Bremen, Germany, operated by the Center of Applied Space Technology and Micro-gravity (ZARM) of the University of Bremen. In this facility, we can achieve up to 9.3 s of low gravity, with a gravity level less than $10^{-4}g$ \citep{ZarmUserManual2023}. This allows us to study the behavior of granular gases in a microgravity environment. The particle dynamics is captured using a high speed camera (EoSens 4CXP from Mikrotron GmbH) with 165 fps. The spatial resolution of our setup makes the mean particle diameter $D$ correspond to 38 pixels.

Fig.~\ref{fig:im1_comp}a shows a raw image from the experiment. Several problems are visible: first, due to space constraints of the magnetic excitation mechanism, the illumination conditions inside the sample cell is inhomogeneous, causing the gray value range of the particle pixels  to  sometimes  overlap with that of the background. In addition, reflections of the surrounding experimental setup are visible on the spherical surface of the sample container.
Finally, a fraction of the particles overlap with each other due to the three-dimensional nature of the experiment.

\subsection{Classical image analysis}
\label{sec:classical}

Classical image segmentation pipelines typically  combine semantic and instance segmentation. The first step is the semantic segmentation where the image is binarized, i.e.,~each pixel is assigned to either the set of all particles or to the background.
Because semantic segmentation is often based on the  pixel gray values,
illumination conditions play an important role at this stage.
The following instance segmentation, where each particle pixel is assigned to an individual particle, becomes an easier task if the relevant pixels have already been identified.  However, the apparent overlap between particles, which is unavoidable for three-dimensional experiments, will complicate this step.

Due to the challenging illumination conditions in our experiment, classical image processing  fails already at the semantic segmentation stage, as shown in figure \ref{fig:im1_comp}(c).  Because of the inhomogeneous illumination
visible in figure \ref{fig:im1_comp}(b),
we could not simply work with one or two global gray value thresholds to identify the pixels belonging to particles; this precludes also the application of more advanced region growing techniques \cite{cheng:20}. We therefore evaluated for each pixel how much its gray value  deviates from that of the background at that same position. Working with the image color instead of the gray value alone did not yield any improvement.

In order to do so we first generated a synthetic background image by analyzing the gray value statistics of each pixel; Figure \ref{fig:im1_comp}(b) demonstrates the success in this task.
Then, if the gray value of an image pixel is within a certain range around the gray value of the same pixel in the background image, we assume the deviation is noise and classify it as a background pixel. Otherwise, we assign that pixel to the set of particle pixels. Even after some postprocessing with morphological filters,  the result in figure \ref{fig:im1_comp}(c) is unsatisfying as many particles appear incomplete or fragmented.

 On the other hand, it is possible for a human observer, making use of geometrical cues and texture information, to identify the actual particles, which suggests that neural networks should be capable of solving this problem. Fig.~\ref{fig:im1_comp}(d) shows that this conclusion is indeed correct and we will explain this method in the following sections.

%===========================================================================================================

\section{Particle detection using an U-Net}
\label{sec:UNet}

\subsection{Creating training, validation, and test data}
\label{sec:data_prep}

Each machine learning project requires a separation of the available data into training data, validation data for the optimization of the  hyper parameters of the model, and test data which provide an independent evaluation of the final model performance. Special care needs to be taken that the test data does not overlap with the training and the validation data since otherwise the performance evaluation would be compromised \citep{kapoor:22}.
During the training of our network, this condition is guaranteed by using separate images for each of the three groups. As shown in Table \ref{tab:training_data}, of the 29 images for which we identify the positions of all particles through human observation, we use 20 for training the network, 7 for validation, and 2 for testing.
\begin{table}[ht]
    \centering
    \begin{tabular}{c c c c}
    Data set    & Number of Images    & Tiles     & Particles\\
    \hline \hline 
    Training    & 20        & 8820      & 9025\\
    \hline 
    Validation  & 7         & 3087      & 3109\\
     \hline
    Testing     & 2         & 882        & 864\\
    %\hline \\
    \hline
    \end{tabular}
    \caption{Sizes of the data sets used for training, validating, and testing.}
    \label{tab:training_data}
\end{table}

\begin{figure}[t]
    \centering
    \includegraphics[width=\columnwidth]{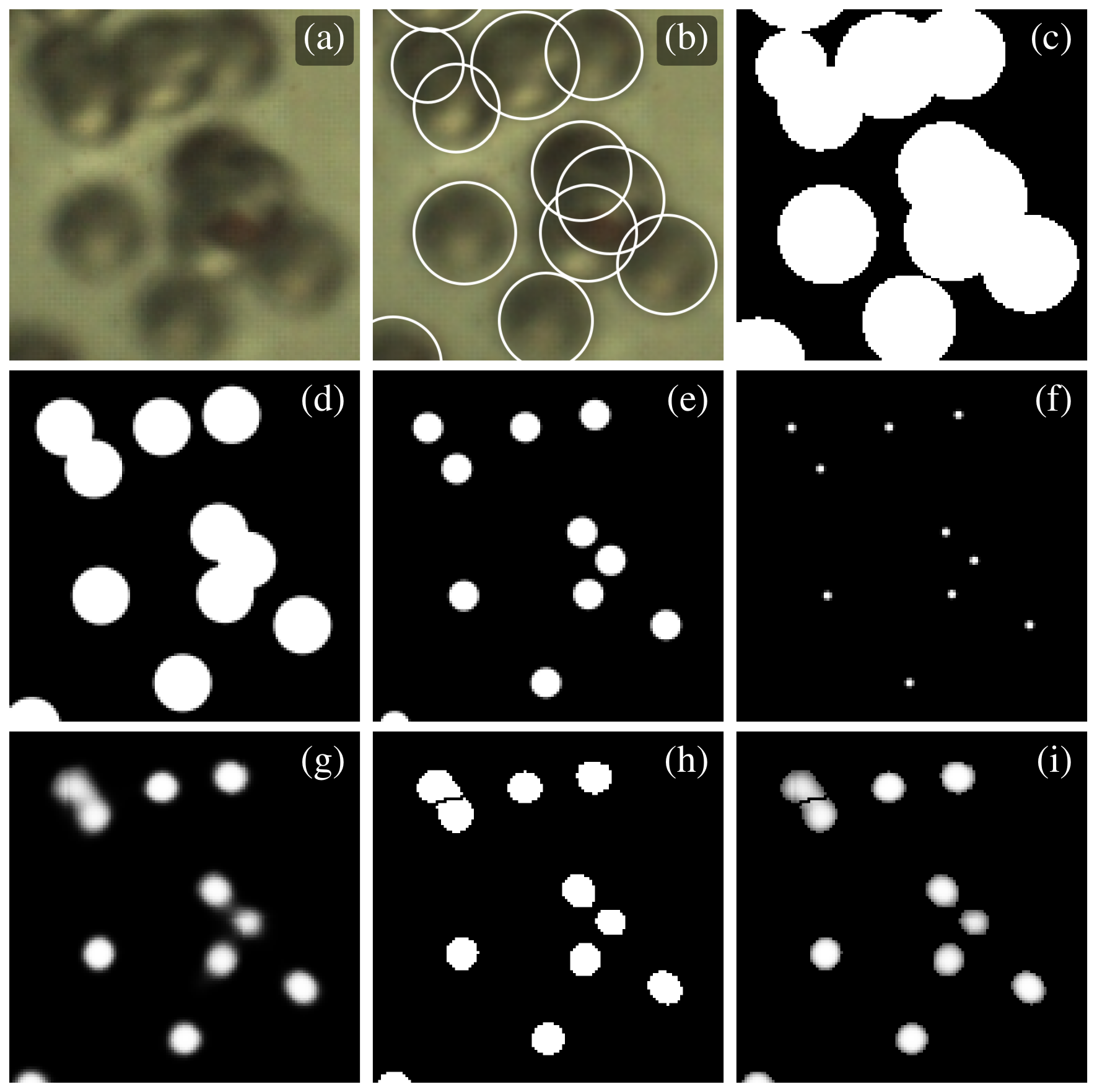}
    \caption{Training of the network requires for each tile a mask image (the target) where the particle positions are indicated with white circular masks:
    \textbf{(a)} Original image tile.
    \textbf{(b)} Manually drawn circles in imageJ used to determine particle coordinates.
    \textbf{(c)} Corresponding mask image with mask radius equal to the particle radius.
    \textbf{(d)-(f)} Mask image with smaller mask radii $R$ = 10, 5, and 1 respectively. Smaller masks help to resolve overlapping particles, as discussed in Section~\ref{sec:mask_size}. 
    \textbf{(g)-(i)} Steps in post-processing: (g) Network output, (h) Water-shed on thresholded binary image and (i) Watershed applied on gray-scale image. Thresholding is done only to removes the noisy pixels.}
    \label{fig:tiles_workflow}
\end{figure}

The raw images taken by our cameras are $1380 \times 1380$ pixels in size, too large for an efficient training of our deep learning architecture. Following \cite{Sanvitale2022}, we cut each image into 441 overlapping tiles of size $128 \times 128$ pixels (see one such tile in Fig.~\ref{fig:tiles_workflow}a.
Adjacent tiles overlap by 50\% of their side length, i.e., by 64 pixels in both horizontal and vertical directions, in order to reduce border artifacts.
Given that each spherical particle has an average diameter of $D \approx 38$ pixels in the images, each tile contains only a few particles. We then manually identify the particle positions in each tile and generate an image tile with masks (see Fig.~\ref{fig:tiles_workflow} c-f) and a list of particle centers. This process is discussed in more details in Section \ref{sec:mask_making}. The full output image is reconstructed by combining the predictions from all tiles while retaining only the central region of each tile to avoid edge effects.

Eventually, 11907 image tiles with size $128\times 128$ are prepared for the training and validation (see table \ref{tab:training_data} for more details).

\subsection{Network architecture and training}
\label{sec:archi_training}
The architecture of a U-Net is described in Fig.~\ref{fig:Unet}. It is trained using an image tile as input and a corresponding manually generated mask tile as ground-truth. U-Nets combine two pathways of information flow. The first one is U-shaped and consist itself of two consecutive steps. First the input image is reduced in size in 4 steps from $128\times 128$ down to  $8\times8$ pixels using max-pooling. At each resolution the image is processed by two consecutive convolution operations combined with an 10 to 30 \% dropout
in order to avoid overfitting \citep{DigitalSreeni2020}. Then the  $8\times8$  feature maps are expanded again to the original size using $2\times2$ up-convolutions. Again, at each size, there are 2 consecutive convolutions with dropout in between. Even though this path partially compensates for the loss of spatial resolution by increasing the number of feature maps from 16 at full resolution to 256 at the $8\times8$ level, the reduction in capacity forces the network to learn the semantic content of the images, quite similar to an autoencoder \cite{hinton:06}.

The second pathway consists of a group of four skip connections which copy, at each resolution, the feature maps from the contraction path to the expansion path (the gray arrows in Fig.~\ref{fig:Unet}). These copies are concatenated with the up-convoluted feature maps so that the subsequent convolutions also have access to the original images at this spatial resolution, which assures a high spatial fidelity of the final result \cite{Unet2015}.

As described in Section \ref{sec:mask_making}, the gray value of each pixel in the mask images describes the fraction of its area covered by a circle of desired radius drawn around the center coordinates. This fixes the value range to 0 (completely outside of any particle) to 1 (completely inside). Therefore, a generic implementation for the output layer is a  $1\times1$ convolution filter with sigmoid activation function acting on the last 16 feature maps.  All other convolution filters use a filter mask of fixed size, between $3\times3$ and  $9\times9$, and Rectified Linear Units (ReLU)  as activation functions. The effect of the filter size will be discussed in Section \ref{sec:filter_size}.
%  We use fractional dropouts with rates of 0.10, 0.10, 0.20, 0.20, and 0.30 at each level of the network to regularize the model.  .....  

\begin{figure}[t]
    \centering
    \includegraphics[width=\columnwidth]{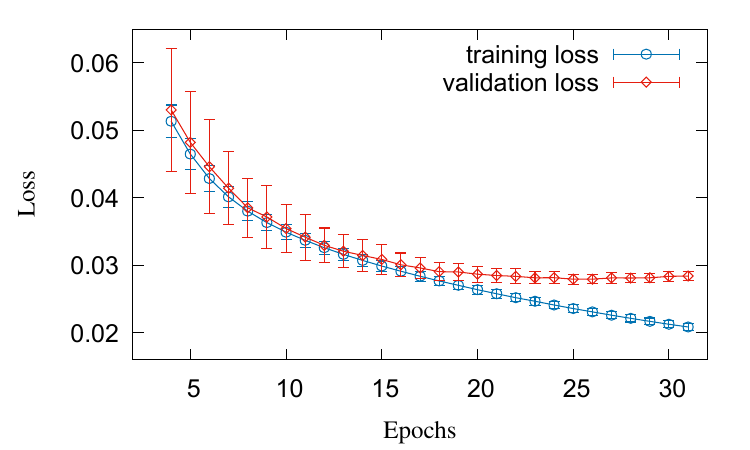}
    \caption{During training, the validation loss typically saturates after 30-60 epochs. Data  is averaged over 10 different random initializations of the same model.}
    \label{fig:epochs_training}
\end{figure}

The network is implemented in TensorFlow using the Keras API. Training is performed with an Adam optimizer and a fixed learning rate of $10^{-4}$ using binary cross entropy as loss. Given the pronounced imbalance between white and black pixels in our masks, we have also tested focal loss \cite{lin:18} and Dice loss~\cite{Sudre2017}. 
We find that differences in the quality of the prediction, as discussed in the next chapter, are well within the standard deviation of those measures. We therefore abide with binary cross entropy.  Figure \ref{fig:epochs_training} shows the evolution of training and validation loss during typical training runs.

%\begin{table}[ht]
%    \centering
%    \begin{tabular}{c c c}
%    Loss    &$F_{\beta}$    &$S_{mean}$\\
%    \hline \hline 
%    DICE    &0.981$\pm$0.002 &1.762$\pm$1.446\\
%    \hline    
%    BFCE    &0.980$\pm$0.002 &1.792$\pm$1.474\\
%    \hline
%    BCE     &0.979$\pm$0.003 &1.798$\pm$1.488\\
%    \hline
% \end{tabular}
%    \caption{\FP{temporary table to compare different losses. For Focal loss (BFCE), 54 different models were trained (6 $\times$9 ($\gamma, \alpha$) combinations), and $\gamma = 2.0, \alpha = 0.60$ was chosen as winner. Concluding, we stick to BCE.}}
% \end{table}

The output image of the U-Net is a gray-scale image where each pixel's value indicates the confidence of the network that this pixel belongs to a particle's mask (which is smaller in diameter than the particle itself). In order to identify the particle centers we use four postprocessing steps in \textit{ImageJ}~\cite{ImageJ2}.
First, the image is binarized with a cutoff value $T$. Section \ref{sec:cut_off}
discusses this hyperparameter in more detail.
Then, the individual particles are identified using the watershed tool\cite{watershed_Beucher1979}  which is internally preceded by an Euclidean Distance transform. This steps also resolves any remaining overlaps. The boundaries found by the watershed algorithm are then used to identify each particle as a list of pixels in the original gray value image (c.f Fig.~\ref{fig:tiles_workflow}(g)-(i)). Finally, the center of mass of each separate list is determined with the
Analyze-Particle tool.

%===========================================================================================================
\subsection{Evaluating the quality of the prediction}
\label{sec:quality}
There are three different aspects which determine the quality of our image processing. First, there is the question if the U-Net does correctly identify all particles visible in the image and also does not hallucinate additional particles. This becomes especially important if we construct particle trajectories using the positions from a sequence of images (not discussed in this manuscript). Second, we care for the accuracy of the positions of the detected particles. This becomes important for determining their precise velocities. Finally, as discussed in Section \ref{sec:classical}, there is the challenge to resolve partially overlapping particles as two entities.

All three of these properties are measured on the validation data in order to learn their dependence on the choice of hyperparameters. Additionally, they are determined relative to the results of human labeling; we do not have access to an independent ground truth.
Here, we first introduce the appropriate measures, the dependency  on the hyperparameters will be discussed in Section \ref{sec:optimization}, the dependence on the human labeler in Section \ref{sec:human_influence}.
\begin{figure}[ht]
    \includegraphics[width=\columnwidth]{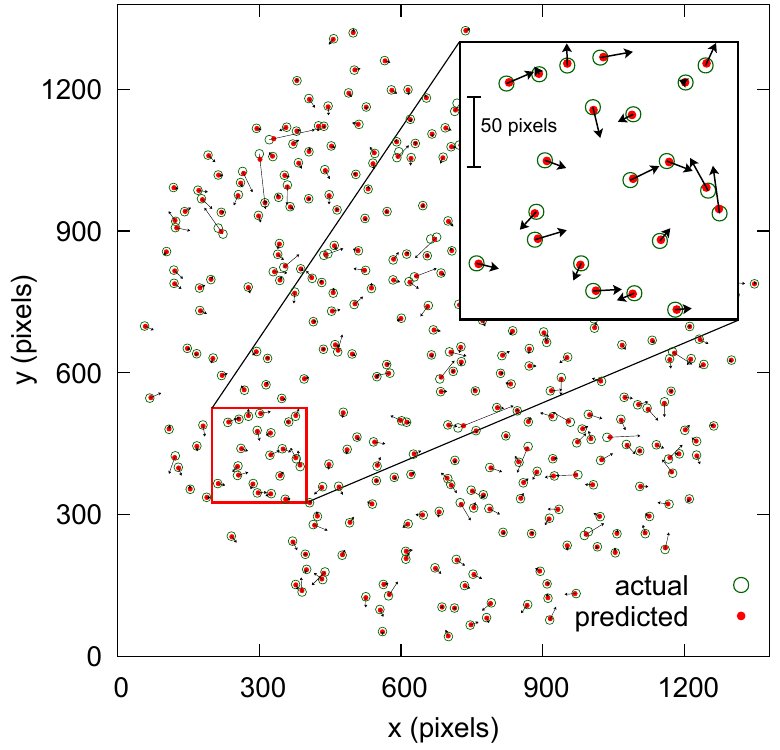}
    \caption{Vectors $\vec{s}$ illustrating the discrepancy between manually identified particle positions (green rings) and the positions determined from the U-Net prediction (red dots). The length $s$ of each vector is magnified 10 times for visibility. 
    }
    \label{fig:sep_vectors}
\end{figure}

For the decision whether a predicted particle position is correct or hallucinated, we find for each predicted particle the nearest particle position in the manually labeled target image. This pair defines a separation vector $\vec{s}$, see Fig. \ref{fig:sep_vectors} for an illustration. 

A particle is considered to be correctly found, or \textit{True Positive} if $|\vec{s}|\leq 20$ pixels\cite{distance_cutoff}. This threshold correspond to about half a particle diameter. If a predicted particle has no partner in the labeled image with $|\vec{s}|\leq 20$ pixels, we consider it a hallucination or \textit{False Positive} ($FP$). Finally, there are \textit{False Negatives} ($FN$), i.e.~manually labeled particles that have no predicted counterpart in a radius smaller than 20 pixels.

Using these three labels, we can compute two quality metrics: the \textit{precision}, $p=TP/(TP+FP)$ and the \textit{recall}, $r=TP/(TP+FN)$ where we would aim for both to be as close to 1 as possible. However, we decided to optimize the hyperparameters of our U-Net using a combined measure, the $F_{\beta}$ score defined as
\begin{equation*}
    F_{\beta} = \frac{(1+\beta^2)\cdot TP}{(1+\beta^2)\cdot TP+FP+\beta^2 \cdot FN}
\end{equation*}
By choosing $\beta = 2$, we give a higher weight to \textit{recall} and thus to minimizing $FN$. This decision is motivated by our aim to subsequently reconstruct the three-dimensional trajectories of the particles. The necessary algorithms are not symmetric with regard to the presence of $FN$ and $FP$: while a $FN$ particle will split a trajectory into two shorter parts, a $FP$ particle will typically be ignored due to the absence of possible predecessor  or successor particles in the time series.

In order to measure the ability of the network to identify partially overlapping particles, we measure  the percentage of particle pairs that are correctly detected as two separate centers rather than a single merged detection. This percentage is a function of the separation distance between the two particle centers; its dependence on mask size will be discussed in Section \ref{sec:mask_size}.

Finally, the separation vector $\vec{s}$ is also used directly to quantify the quality of the predictions: optimally it would be as small as possible for all particles in the validation dataset. Additionally, any measurable deviation from an isotropic angular distribution points to a systematic error.

%%%%%%%%%%%%%%%%%%%%%%%%%%%%%%%%%%%%%%%%%%%%%%%%%%%%%%%%%%%%%%%%%%%%%%%%%%%%%%%%%%%%%%%%
%%%%%%%%%%%%%%%%%%%%%%%%%%%%%%%%%%%%%%%%%%%%%%%%%%%%%%%%%%%%%%%%%%%%%%%%%%%%%%%%%%%%%%

\section{Creating mask images for training}
\label{sec:create_mask}
The U-Net results can only be as good as the image and mask pairs it has been trained with. Therefore, the preparation of high quality masks is paramount. Creating mask images requires two consecutive steps: First a human labeler has to create a list of the particle centers, then the mask image has to be created by drawing filled white circles at those positions into an otherwise black image. In Sections \ref{sec:imageJ} and \ref{sec:mask_making} we will discuss some caveats regarding these two steps.  Section \ref{sec:human_influence} describes how different human
labeler have different biases and how an ensemble of labelers can be used to finetune the model.

%%%%%%%%%%%%%%%%%%%%%%%%%%%%55
\subsection{Identifying particle centers}
\label{sec:imageJ}
While there are other open source solutions for labeling training data \cite{LabelStudio}, we found the image processing program \textit{ImageJ}~\cite{ImageJ2} most suitable for this task.
After the image is loaded, the human labeler can zoom into the image as desired and draw
circles around the identifiable particles, as shown in Fig.~\ref{fig:tiles_workflow}b.  \textit{ImageJ} processes this information and outputs a list of the particle centers as floating point values. A set of instructions for training prospective labeler is included with the code and data on github\cite{github}.

\begin{figure}[ht]
    \centering
    \includegraphics[width=0.90\columnwidth]{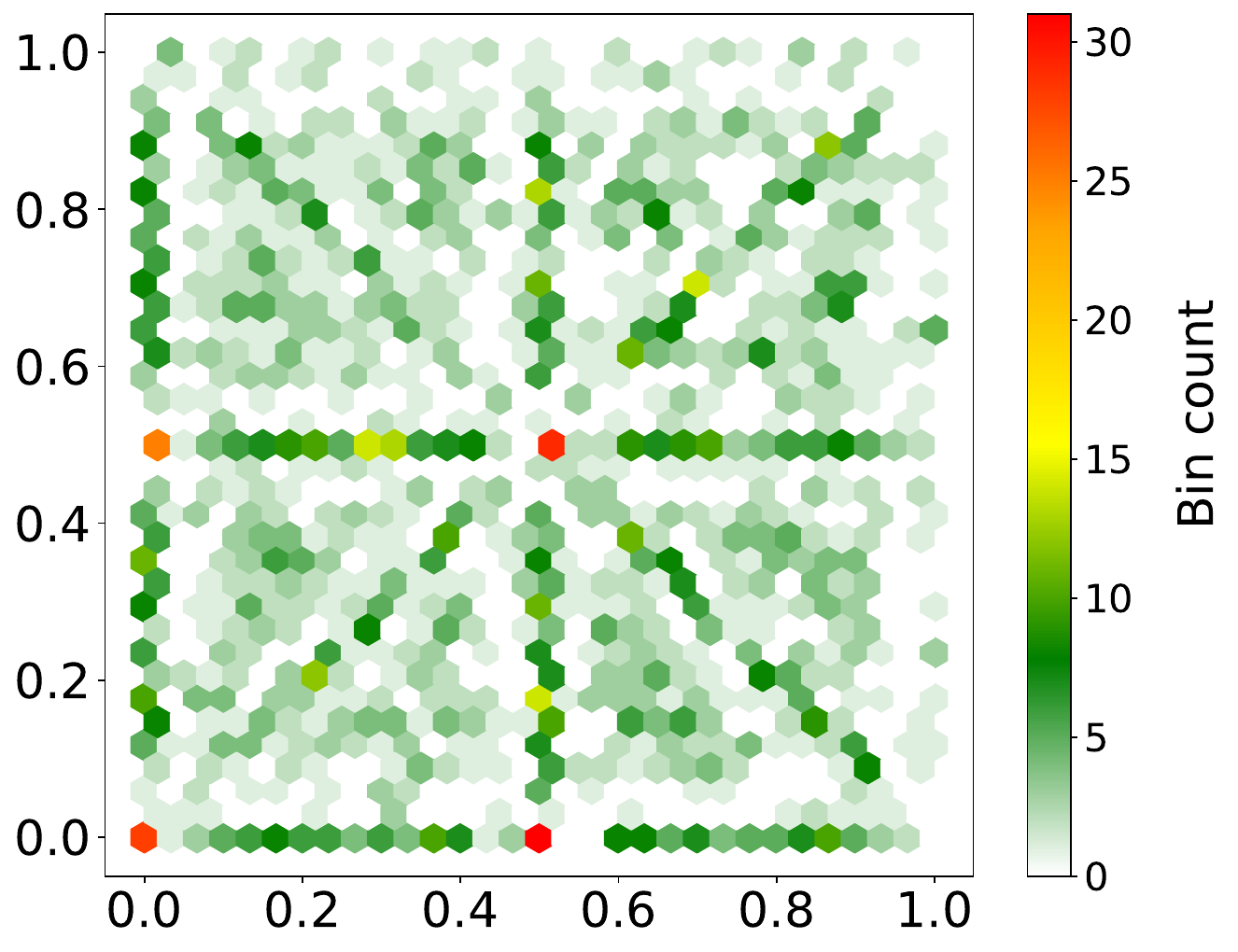}
    \caption{The distribution of decimal places in the particle coordinates reported by ImgageJ is not uniform. The color bar shows the frequency of coordinate pairs falling in each hexagonal bin, which reveals a bias towards integer or half-integer values. The data is generated from 2157 particle coordinates obtained from 5  images.
    } 
    \label{fig:decimal_distr}
\end{figure}

During the labeling of our images we found that ImageJ has a minor issue. One would expect the decimal places of the particle coordinates to be uniformly distributed between 0 and 1 in both the x and y directions.  However, as shown in Figure~\ref{fig:decimal_distr}, this is not the case: the likelihood of obtaining an integer or a half-integer value is significantly higher than the average. This behavior can be traced back to a form of 'snapping' of the coordinate values; it can be reproduced by drawing slightly shifted circles at a high magnification.  Based on the insight that the imperfections of the human labelers (discussed in section \ref{sec:human_influence}) are larger than the error created by ImageJ, we decided to accept this limitation.

%%%%%%%%%%%%%%%%%%%%%%%%%%%%%%%%
\subsection{Making binary mask images}
\label{sec:mask_making}

The second step in the creation of the mask images is to use the particle coordinates to draw filled white circles of radius $R$ in an otherwise black image. Figure~\ref{fig:tiles_workflow} (d) to (f) gives an example. The choice of $R$ is an important hyperparameter; its influence on the results is discussed in 
Section \ref{sec:mask_size}.

A naive way of doing so would be to round the particle coordinates to the next integer value and use the corresponding pixel as the center of the circle to be drawn. However, this would introduce a systematic error due to the difference between the actual pixel center and the coordinates used to address it. For example, let us assume our center coordinates are $x$ = 12.4 and $y$ = 12.6.  If our mask had the size of only one pixel, we should choose the pixel indexed (12,12) which has a center of 12.5, 12.5 in the global coordinate system  used to measure particle centers. However, rounding to the next integer results in the pixel indexed (12,13) being chosen as the center.

\begin{figure}[t]
    \centering
    \includegraphics[width=\columnwidth]{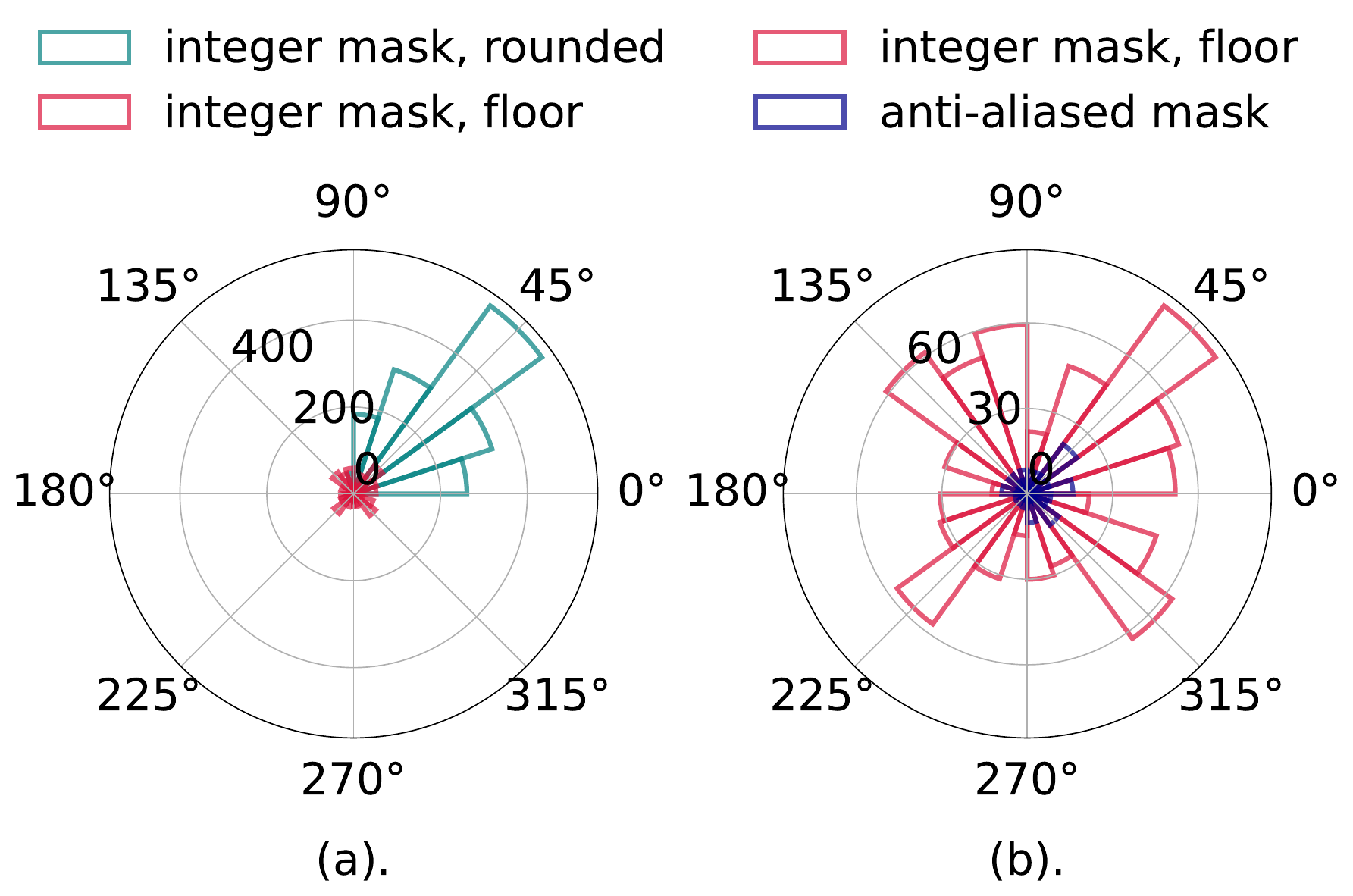}
    \caption{A tool to study systematic bias are polar histograms of displacement vectors between particle centers obtained during manual labeling and those extracted from the corresponding mask images. The vectors are grouped into 20 angular bins of 18$\degree$ each, and the radial value indicates the sum of vector magnitudes within each bin. Histograms are computed from 2157 particle coordinates.
    \textbf{(a)} Integer mask centers obtained by rounding introduce a systematic bias, whereas using the floor operation removes this effect.
    \textbf{(b)} Anti-aliased, floating-point-based masks further reduce the positional error. For visibility, the displacement magnitudes of anti-aliased masks in (b) are scaled by a factor of ten.
    }
    \label{fig:comparison}
\end{figure}

The correct way of turning the floating point numbers into integers is therefore the floor operation, because it assures that all decimal values betweeen 0.000... and 0.9999... get mapped onto the integer below, which then creates a particle center at 0.5.
This point is also proven numerically in figure~\ref{fig:comparison} (a).

A second, more subtle issue of integer based mask centers arises from the uneven distribution shown in figure \ref{fig:decimal_distr}.  The red-colored maxima
at $x$ and/or $y$ equals zero show that real centers in some unknown range around zero will snap to exactly zero as decimal place.
For centers located slightly left and/or below those red points, this results in an erroneous increase of the center position by one pixel.
For example, a real particles center of 11.98 which would be reported as 12 by ImageJ, leading to a mask center of 12.5 instead of 11.5 \cite{random_perturbation}.

% A general solution of the problems of these integer based masks is to eliminate the rounding process altogether and create anti-aliased masks with centers based on the floating point value from the identification stage. This approach is discussed in the next section.

\subsubsection{Anti-aliased masks}
\begin{figure}[t]
    \centering
    \includegraphics[width=\columnwidth]{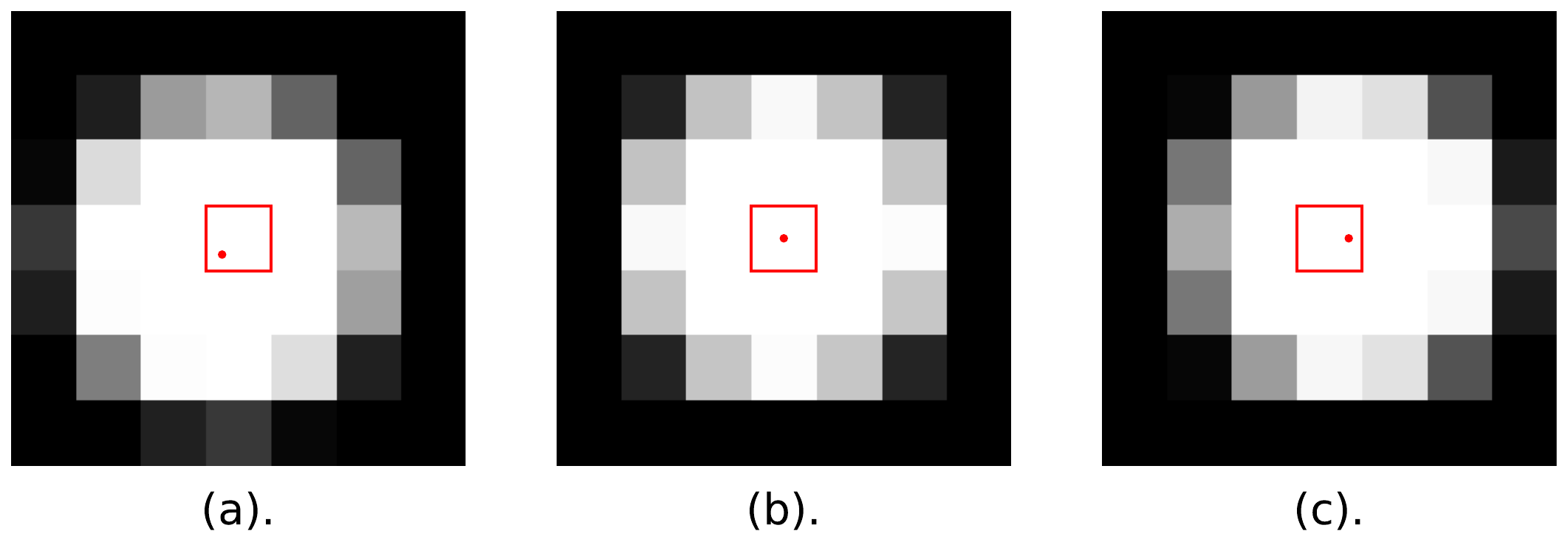}
    \caption{Anti-aliased masks can be centered around floating point positions (the red dots). Shown are masks with a radius of 2 pixels centered at (a) $x=3.25, y=3.25$, (b) at $x=3.50, y=3.50$ and (c) at $x = 3.80, y = 3.50$. The red square indicates the integer center of the masks which is identical in all three cases.}
    \label{fig:anti_ali_pos}
\end{figure}

The best way to avoid all the problems associated with an integer based center position of the particle masks is to use anti-aliased masks which can be positioned with floating point accuracy. To create these masks, an imaginary circle of the intended radius is overlaid over a grid of square pixels. 
All pixels which are completely within the circle are set to white. Pixels that are cut into two by the circle are colored with a gray value which is proportional to the pixel area within the circle.  Figure~\ref{fig:anti_ali_pos}
gives an example how these anti-aliased masks change with the exact position of the mask center.
Figure \ref{fig:comparison} (b) depicts the clear superiority of these 
anti-aliased masks over the integer based ones.

%%%%%%%%%%%%%%%%%%%%%%%%%555

\subsection{The influence of the human labeler }
\label{sec:human_influence}
To investigate the influence of human labeler on the ground truth coordinates, we conducted an experiment where multiple people were asked to manually annotate the same images. The coordinates obtained from each labeler were then grouped using the DBSCAN algorithm \cite{ester:96} to identify particles that were consistently labeled by all participants. For each particle, we calculated the mean coordinate and the separation of each labeler's coordinate from this mean position. By averaging these separations across all particles identified by each labeler, we obtained a mean separation vector for each labeler.

\begin{figure}[]
    \centering
    \includegraphics[width=\columnwidth]{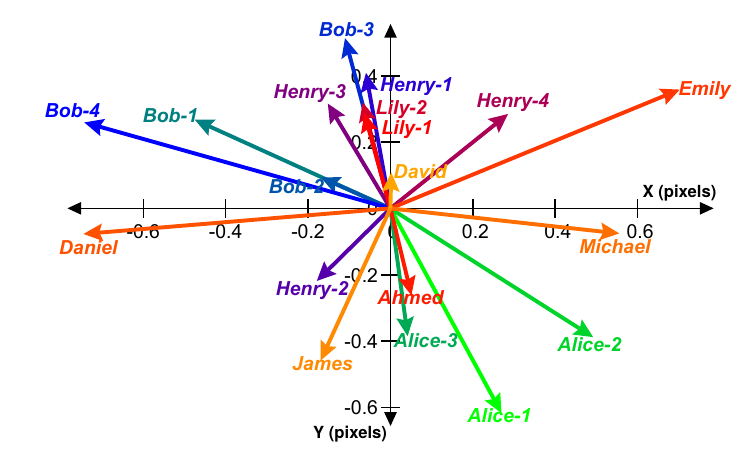}
    \caption{Human labeler exhibit a systematic bias in their annotations. Each vector represents the mean deviation of a labeler's coordinates for one image from the consensus (mean) positions using all labeler's coordinates. Alice, Bob, Henry, and Lily vectors have labeled multiple images. While Alice, Bob, and Lily exhibit consistent bias across images, Henry’s deviations vary significantly.}
    \label{fig:sep_vecs_everyone}
\end{figure}

Figure \ref{fig:sep_vecs_everyone} shows the mean separation vectors for each labeler. Some labelers show a clear directional bias. E.g.~Alice’s annotations across three images consistently deviate into the 4th quadrant, while Bob’s four annotations all lie in the 2nd quadrant.
In contrast, Henry’s deviations vary significantly across images.

\begin{figure*}[t]
    \centering
    \includegraphics[width=1.60\columnwidth]{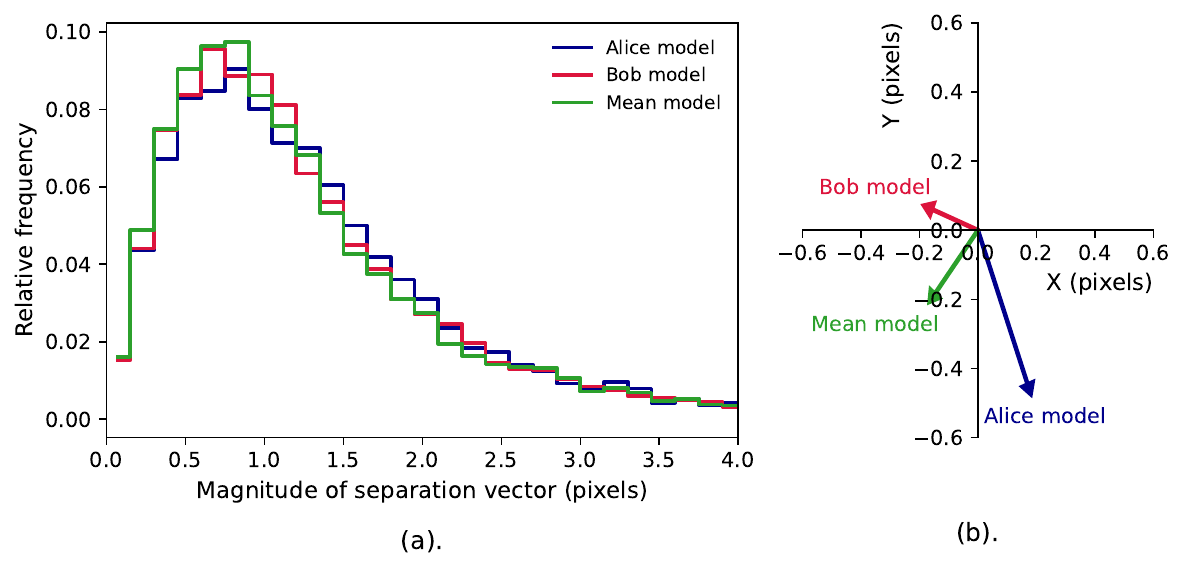}
    \caption{Fine-tuning the model using the mean coordinates from multiple labelers as ground truth reduces bias. These graphs are based on the displacement vectors $\vec{s}$ for the two test images, each labeled by eight labelers. The Alice and Bob models are the models fine-tuned using 13 and 7 images labeled solely by Alice and Bob, respectively, whereas the Mean model is the model fine-tuned using the mean coordinates from 10 labelers for two training images.
    (a) distribution of magnitudes of the separation vectors $\vec{s}$ shifts toward lower values after fine-tuning, resulting in a smaller mean separation magnitude. (b) The total bias can be characterized by the mean displacement vector which is computed by averaging $\vec{s}$ over all directions. Bob model and Alice model consistently reproduces their respective biases identified from separation of their labels from mean coordinates as shown in Fig.~\ref{fig:sep_vecs_everyone}.}
    \label{fig:fine_tuned_separation}
\end{figure*}

The magnitudes of these vectors are relatively small, ranging upto 0.7 pixels in both x and y components, suggesting that while the biases are real, they are also relatively subtle. However, the network learns these biases: if the U-Net is trained using only the 13 training images labeled by Alice or the 7 training images labeled by Bob it displays a consistent directional bias in its predictions, as shown in Fig.~\ref{fig:fine_tuned_separation}(b). This sensitivity to training biases needs to be considered when all the training data are created by a single person only.

 By acknowledging and accounting for these biases, we can develop a more robust and reliable image segmentation algorithm. The initial training of our model used as ground only the 20 images analyzed by Alice and Bob.
We additionally collected labels on 2 images from 10 different labelers each and computed the mean positions of the particles they found. Then we fine-tuned the initial model by retraining it using those mean positions; we call this final model the mean model.

For the final evaluation we had each 8 labelers analyzing our two test images. Figure \ref{fig:fine_tuned_separation}(a) shows the distribution of the  magnitude of separation vectors calculated from mean coordinates of the two test images. The average magnitude of the separation vectors is 1.35 pixel for the mean model; the models trained on Bob's of Alice images alone are slightly worse with  $1.44$ to $1.38$ pixels. So our final best result for the mean deviation of the U-Net prediction from the best estimate of the ground truth (the averaged labels) is  $1.35$ pixels corresponding to 3.6\% of a particle diameter.

%%%%%%%%%%%%%%%%%%%%%%%%%%%%%%%%%%%%%%%%%%%%%%%%%%%%%%%%%%%%%%%%%%%%%%%%%%%%%%%%%%%%%%%
%%%%%%%%%%%%%%%%%%%%%%%%%%%%%%%%%%%%%%%%%%%%%%%%%%%%%%%%%%%%%%%%%%%%%%%%%%%%%%%%%%%%%%%

\section{Optimizing the hyperparameters}
\label{sec:optimization}

As discussed in Section \ref{sec:quality}, we want to optimize three different metrics with our algorithm: the $F_2$ score for minimizing the False Positives and False Negatives, the mean separation vector $\vec{s}$ for minimizing the difference between predicted position and ground truth, and the percentage of correctly identified overlapping sphere pairs which we want to maximize. The most important hyperparameters to reach this objective are the radius of our masks $R$, the threshold $T$ used for binarizing the prediction of our U-Net and the filter size $f$ used in the convolutions

\subsection{Optimizing the detection of overlapping particles}
\label{sec:mask_size}
As discussed in Section \ref{sec:classical}, partially overlapping particles pose a challenge for any segmentation algorithm. It is intuitive that the U-Net has a better chance of distinguishing them when the masks representing the two particles do not overlap. (Note that even if two particles in the output overlap, the watershed step described in Section \ref{sec:archi_training} still allow us to separate them.)

\begin{figure}[t]
    \centering
    \includegraphics[width=\columnwidth]{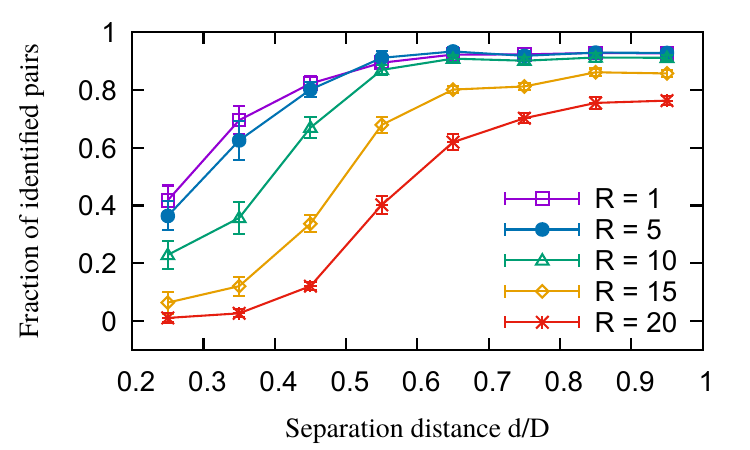}
    \caption{The fraction of correctly identified pairs of particles improves with decreasing mask size $R$ for separation distances $d$ that are smaller than the particle diameter $D$ of 38 pixels. Data is averaged over 7 validation images, with 10 different network initializations used to generate the error bars. All measurements are done with $T$ values maximizing the $F_2$ and filter size $f$=3.}
    \label{fig:delta_distribution}
\end{figure}

This implies that the radius of the masks  $R$  should be significantly smaller than half the particle diameter $D/2$, i.e.~smaller than 19 pixels for our images. 
Fig. \ref{fig:delta_distribution} shows the capability of U-Nets trained with different mask sizes to handle overlapping particles. It can be seen that $R$ values smaller than 5 do not result in an improvement. As shown below, $R$ influences also our other target metrics which therefore also have to be taken into account when choosing an optimal value.
 
\subsection{Minimizing the mean positional error $\vec{s}$}
\begin{figure}[ht]
    \centering
    \includegraphics[width=\columnwidth]{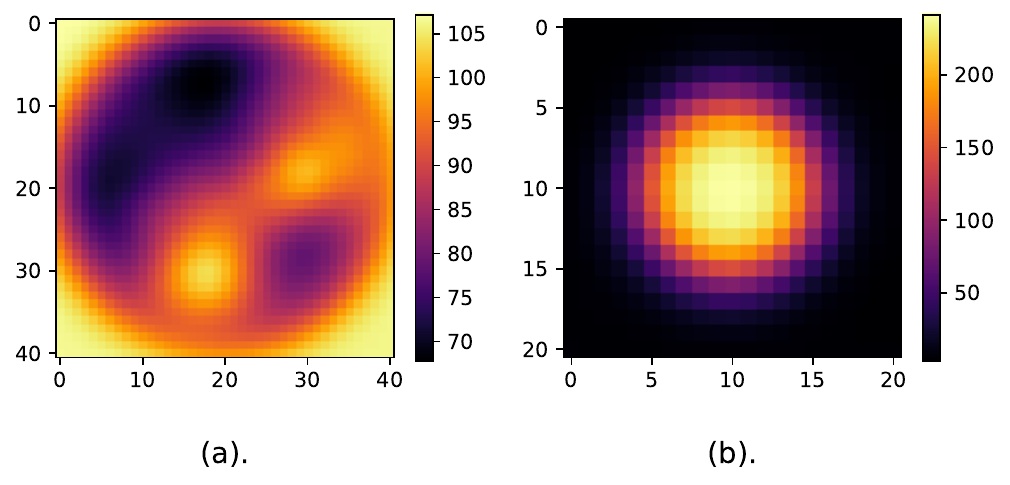}
    \caption{Even though the particle images are irregular, the U-Net learns to represent the particles as radially symmetric. (a) Knowing the particle centers, we can average the individual particles images. The result demonstrates the inhomogeneous illumination of our sample. (b) the average over the predicted outputs (here with $R$ = 5 pixels) shows no asymmetry. 
     %\color{red} ((a) is only red channel.)
     }
    \label{fig:particle_tiles}
\end{figure}

\begin{figure}[ht]
    \centering
    \includegraphics[width=\columnwidth]{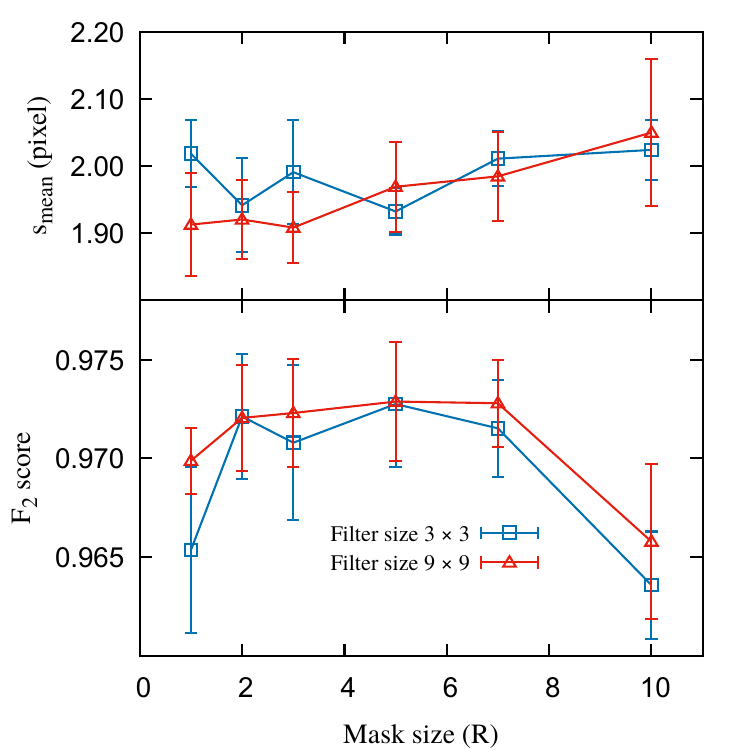}
    \caption{Given the size of the uncertainty bars, a mask radius $R$ = 5 and 
    a filter size $f$ = 3 are good choices to maximize the $F_2$ score and at the same time 
    achieve a small mean separation $s_{mean}$. 
    Data is obtained by averaging the results from 10 models trained with different random initializations and evaluated on 7 validation images. $T$ values were chosen to optimize the $F_2$ score. Uncertainty bars are standard deviations. Results are measured using the validation data and the model trained on Alice's and Bob's labels only.
    }
    \label{fig:mask-filter_size_comp}
\end{figure}

Given that the anti-aliased masks we use for training (c.f.~figure \ref{fig:anti_ali_pos}) cover at their boundaries
the full range of gray values, it is unsurprising that the predicted particle position are also not binary images. However, as figure \ref{fig:particle_tiles} (b)
shows, the gray value boundaries of the predicted particles extend over a larger range of radii. In order to obtain a precise location we therefore binarize the predicted images with a threshold $T$ as described in Section \ref{sec:archi_training}.  Since the mean output shown in figure \ref{fig:particle_tiles} (b) is symmetric, we can expect $T$ to not influence the size or direction of the individual $\vec{s}$ vectors. In contrast, the mask radius and the convolutional filter size do have a weak influence as shown in the top part of figure \ref{fig:mask-filter_size_comp}. 

\subsection{Maximizing the $F_2$ score}
\label{sec:cut_off}
\label{sec:filter_size}
\begin{figure}[t]
    \centering
    \includegraphics[width=\columnwidth]{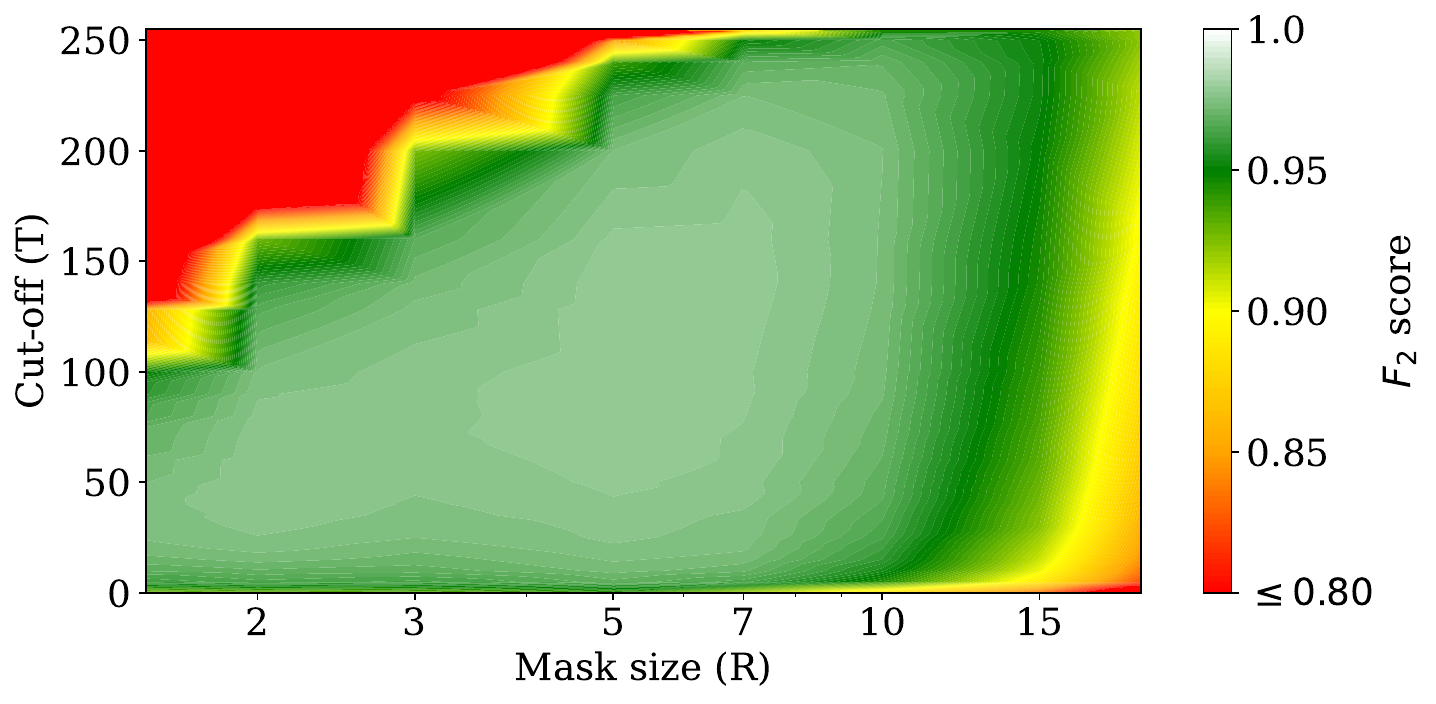}
    \caption{For all values of $R$ there is intermediate value of the cut-off $T$ which maximizes the $F_2$ score. All results were obtained with a convolution filter size of 3.
    }
    \label{fig:cutoff_dep}
\end{figure}

The cut-off threshold $T$ determines which part of the U-Net output is included in the analysis. Setting it too low will discard particles which did not create a strong output signal, setting it too high will merge the gray value clouds of neighboring particles. Fig.~\ref{fig:cutoff_dep} shows how $T$ and $R$ influence the $F_2$ score. For $R$ = 5 we obtain a broad maximum, allowing us to optimize all three of our metrics at the same time.
The bottom part of Fig.~\ref{fig:mask-filter_size_comp} shows that this maximum is also not depending on $f$.

\subsection{Final results based on test image}
Using a set of two test images where the coordinates have been averaged over $8$ labelers for each, we find our U-Net fine-tuned on mean coordinates(see Section \ref{sec:human_influence}) has an $F_2$ score of $0.973$. This corresponds to correctly identifying 97.5 \% of the particles while only creating 3.5 \% of false positives. The mean accuracy of the particle coordinates is 3.6 \% of their diameter.

\section{Conclusion}
\label{sec:conclusion}
We have shown that even challenging images of granular experiments can be successfully segmented using a U-Net architecture. We also demonstrated how paying attention to the details in the generation of the training data, including the biases introduced by human labelers, is important to assure optimal results. The optimized U-Net will provide the basis of our future particle tracking work.

A potential future optimization is the adaptation of the current U-Net to the detection of granular particles with irregular shapes. Existing applications \cite{liu:20,wang:21,yu:25} in rock fragments segmentation have already shown the capability of U-Net in this aspect. The relative insensitivity of our results to the hyperparameters indicates that all information contained in the images has been successfully extracted. A way of proving that point would be a comparison with alternative deep-learning based segmentation algorithms. As we provide both our trained model and the training, validation, and test data under an open source license, this work could serve as a benchmark.

\backmatter

\bmhead{Acknowledgments}
We are especially grateful to Sophia Grille for her extensive contribution to the dataset annotation, contributing the majority of the annotations used for the initial U-Net training. We thank also thank our additional labelers, Kai Albert, Juan Carlos Araño Romero, Andrii Demydenko, Leonardo Facchini, Dorian Feisel, Flynn Frank-Richter, Mark Goh, Ansab Moosa, L\'ea Gommeringer, Sofie H\"ubl, Jiahe Yu, Yona Lee, Akash Mangala, Marimel Mayer, Sibylle N\"agele, Nuria Navarrete Argiles, Yasmin Schr\"oter, Nisha Singh, Juri Th\"ummler, and Golo Zimmermann for their time. We also thank Dmitry Puzyrev, Karsten Tell, and the AIMS team for helpful discussions.
\section*{Declarations}

\textbf{Funding}
We acknowledge partial funding support from DLR through Project No. 50WK2270D (Artificial Intelligence for Granular Experiments, AIGE, a sub-project of Artificial Intelligence Meets Space, AIMS). Fahad Puthalath is sup-
ported by Hanns-Seidel-Stiftung (HSS) and by
German Academic Exchange Service (DAAD). The authors gratefully acknowledge the computational and data resources provided through the joint high-performance data analytics (HPDA) project “terrabyte” of the German Aerospace Center (DLR) and the Leibniz Supercomputing Center (LRZ).

\textbf{Conflict of interest} The authors declare no Conflict of interest. 

\textbf{Code availability} 
The code of the U-Net, the weights, all training, validation, and test images including labels, and instructions for aspiring labelers will be published on GitHub\cite{github} under
CC BY-SA 4.0 license \cite{cc}.

\bibliography{ref}% common bib file

@misc{github,
  author = {Fahad Puthalath},
  title = {UNet-for-Particle-detection},
  year = 2026,
  howpublished = {\url{https://github.com/pfahad/UNet-for-Particle-detection}},
}

@article{wang:21,
title = {An {Improved} {Boundary}-{Aware} {U}-{Net} for {Ore} {Image} {Semantic} {Segmentation}},
	volume = {21},
	copyright = {http://creativecommons.org/licenses/by/3.0/},
	doi = {10.3390/s21082615},
	journal = {Sensors},
	publisher = {Multidisciplinary Digital Publishing Institute},
	author = {Wang, Wei and Li, Qing and Xiao, Chengyong and Zhang, Dezheng and Miao, Lei and Wang, Li},
	year = {2021},
	pages = {2615},
}

@article{liu:20,
	title = {Ore image segmentation method using {U}-{Net} and {Res}\_Unet convolutional networks},
	volume = {10},
	doi = {10.1039/C9RA05877J},
	journal = {RSC Advances},
	publisher = {The Royal Society of Chemistry},
	author = {Liu, Xiaobo and Zhang, Yuwei and Jing, Hongdi and Wang, Liancheng and Zhao, Sheng},
	year = {2020},
	pages = {9396--9406},
}

@article{cheng:18,
	title = {A particle-tracking method for experimental investigation of kinematics of sand particles under triaxial compression},
	volume = {328},
	doi = {10.1016/j.powtec.2017.12.071},
	journal = {Powder Technology},
	author = {Cheng, Zhuang and Wang, Jianfeng},
	year = {2018},
	pages = {436--451},
}

@article{cheng:21,
	title = {An investigation of the breakage behaviour of a pre-crushed carbonate sand under shear using {X}-ray micro-tomography},
	volume = {293},
	doi = {10.1016/j.enggeo.2021.106286},
	journal = {Engineering Geology},
	author = {Cheng, Zhuang and Wang, Jianfeng},
	year = {2021},
	pages = {106286},
}

@article{cheng:20_b,
	title = {Tracking particles in sands based on particle shape parameters},
	volume = {31},
	doi = {10.1016/j.apt.2020.02.033},
	journal = {Advanced Powder Technology},
	author = {Cheng, Zhuang and Zhou, Bo and Wang, Jianfeng},
	year = {2020},
	pages = {2005--2019},
}

@article{cheng:20,
	title = {Improved region growing method for image segmentation of three-phase materials},
	volume = {368},
	doi = {10.1016/j.powtec.2020.04.032},
	journal = {Powder Technology},
	author = {Cheng, Zhuang and Wang, Jianfeng},
	year = {2020},
	pages = {80--89},
}

@article{yu:25,
	title = {A method for identifying fragmentation of open-pit mining blasting based on a new hybrid convolutional neural network},
	volume = {27},
	doi = {10.1007/s10035-025-01542-7},
	journal = {Granular Matter},
	author = {Yu, Jianyang and Meng, Lingyu and Ren, Shijie and Song, Xubin and Liang, Hongzhi and Cao, Jiachen and Xue, Yanping and Zhou, Wangbin},
	year = {2025},
	pages = {62},
}

@misc{cc,
    note = {https://creativecommons.org/licenses/by-sa/4.0/}
}

@INPROCEEDINGS{ester:96,
  author={Ester, Martin and  Kriegel, Hans-Peter and Sander, J\"org and  Xu, Xiaowei },
  booktitle={Proceedings of the Second International Conference on Knowledge Discovery and Data Mining (KDD-96)}, 
  title={A density-based algorithm for discovering clusters in large spatial databases with noise}, 
  year={1996},
  pages={226–231},
}

@misc{random_perturbation,
    note = {This effect could be counteracted by adding a small random perturbation $-0.01<\epsilon<0.01$ to every  particle position that happens to be an integer. Together with rounding by floor, this would with a 50\% probability change the position of those particles coming from the 'snapping region' to the next smaller integer. However, a better way to solve the problem the are anti-aliased masks described in the next subsection.} 
}

@misc{distance_cutoff,
  note={The F2 score depends on the chosen distance threshold; however, it saturates for thresholds above approximately 8 pixels, as increasing the cutoff does not yield additional correct matches.}
}

@misc{kapoor:22,
	title = {Leakage and the {Reproducibility} {Crisis} in {ML}-based {Science}},
	doi = {10.48550/arXiv.2207.07048},
	publisher = {arXiv},
	author = {Kapoor, Sayash and Narayanan, Arvind},
	year = {2022},
	note = {arXiv:2207.07048 [cs, stat]},
}

@InProceedings{Unet2015,
author="Ronneberger, Olaf
and Fischer, Philipp
and Brox, Thomas",
editor="Navab, Nassir
and Hornegger, Joachim
and Wells, William M.
and Frangi, Alejandro F.",
title="U-Net: Convolutional Networks for Biomedical Image Segmentation",
booktitle="Medical Image Computing and Computer-Assisted Intervention -- MICCAI 2015",
year="2015",
publisher="Springer International Publishing",
address="Cham",
pages="234--241",
isbn="978-3-319-24574-4",
doi={10.1007/978-3-319-24574-4_28}
}

@article{Liu2022,
  doi = {10.1142/s1793545822500316},
  year = {2022},
  publisher = {World Scientific Pub Co Pte Ltd},
  volume = {15},
  author = {Zhichao Liu and Heng Zhang and Luhong Jin and Jincheng Chen and Alexander Nedzved and Sergey Ablameyko and Qing Ma and Jiahui Yu and Yingke Xu},
  title = {U-Net-based deep learning for tracking and quantitative analysis of intracellular vesicles in time-lapse microscopy images},
  journal = {Journal of Innovative Optical Health Sciences}
}

@article{Newby2018,
author = {Jay M. Newby  and Alison M. Schaefer  and Phoebe T. Lee  and M. Gregory Forest  and Samuel K. Lai },
title = {Convolutional neural networks automate detection for tracking of submicron-scale particles in {}2D and {3D}},
journal = {Proceedings of the National Academy of Sciences},
volume = {115},
pages = {9026-9031},
year = {2018},
doi = {10.1073/pnas.1804420115}
}

@article{Midtvedt2021,
    author = {Midtvedt, Benjamin and Helgadottir, Saga and Argun, Aykut and Pineda, Jesús and Midtvedt, Daniel and Volpe, Giovanni},
    title = "{Quantitative digital microscopy with deep learning}",
    journal = {Applied Physics Reviews},
    volume = {8},
    year = {2021},
    doi = {10.1063/5.0034891},
    note = {011310},
}

@Article{Puzyrev2020,
author={Puzyrev, Dmitry
and Harth, Kirsten
and Trittel, Torsten
and Stannarius, Ralf},
title={Machine Learning for 3D Particle Tracking in Granular Gases},
journal={Microgravity Science and Technology},
year={2020},
volume={32},
pages={897-906},
doi={10.1007/s12217-020-09800-4},
}

@Article{Sanvitale2022,
author={Sanvitale, Nicoletta
and Gheller, Claudio
and Bowman, Elisabeth},
title={Deep learning assisted particle identification in photoelastic images of granular flows},
journal={Granular Matter},
year={2022},
volume={24},
pages={65},
doi={10.1007/s10035-022-01222-w},
}

@article{Yu2020,
  title = {Velocity Distribution of a Homogeneously Cooling Granular Gas},
  author = {Yu, Peidong and Schr\"oter, Matthias and Sperl, Matthias},
  journal = {Phys. Rev. Lett.},
  volume = {124},
  issue = {20},
  pages = {208007},
  numpages = {6},
  year = {2020},
  month = {May},
  publisher = {American Physical Society},
  doi = {10.1103/PhysRevLett.124.208007},
  url = {https://link.aps.org/doi/10.1103/PhysRevLett.124.208007}
}

@article{Sack2013,
  title = {Energy Dissipation in Driven Granular Matter in the Absence of Gravity},
  author = {Sack, Achim and Heckel, Michael and Kollmer, Jonathan E. and Zimber, Fabian and P\"oschel, Thorsten},
  journal = {Phys. Rev. Lett.},
  volume = {111},
  issue = {1},
  pages = {018001},
  numpages = {5},
  year = {2013},
  month = {Jul},
  publisher = {American Physical Society},
  doi = {10.1103/PhysRevLett.111.018001},
  url = {https://link.aps.org/doi/10.1103/PhysRevLett.111.018001}
}

@Article{Pitikaris2022,
author={Pitikaris, Sebastian
and Bartz, Patricia
and Yu, Peidong
and Cristoforetti, Samantha
and Sperl, Matthias},
title={Granular cooling of ellipsoidal particles in microgravity},
journal={npj Microgravity},
year={2022},
month={Apr},
day={20},
volume={8},
number={1},
pages={11},
issn={2373-8065},
doi={10.1038/s41526-022-00196-6},
url={https://doi.org/10.1038/s41526-022-00196-6}
}

@ARTICLE{Schneider2021,
       author = {{Schneider}, Niclas and {Musiolik}, Grzegorz and {Kollmer}, Jonathan E. and {Steinpilz}, Tobias and {Kruss}, Maximilian and {Jungmann}, Felix and {Demirci}, Tunahan and {Teiser}, Jens and {Wurm}, Gerhard},
        title = "{Experimental study of clusters in dense granular gas and implications for the particle stopping time in protoplanetary disks}",
      journal = {Icarus},
         year = 2021,
        month = may,
       volume = {360},
          eid = {114307},
        pages = {114307},
          doi = {10.1016/j.icarus.2021.114307}
}

@article{Falcon1999,
  title = {Cluster Formation in a Granular Medium Fluidized by Vibrations in Low Gravity},
  author = {Falcon, \'Eric and Wunenburger, R\'egis and \'Evesque, Pierre and Fauve, St\'ephan and Chabot, Carole and Garrabos, Yves and Beysens, Daniel},
  journal = {Phys. Rev. Lett.},
  volume = {83},
  issue = {2},
  pages = {440--443},
  numpages = {0},
  year = {1999},
  month = {Jul},
  publisher = {American Physical Society},
  doi = {10.1103/PhysRevLett.83.440},
  url = {https://link.aps.org/doi/10.1103/PhysRevLett.83.440}
}

@Article{Yu2019,
    author = {Yu, Peidong and Stärk, Elmar and Blochberger, Guido and Kaplik, Martin and Offermann, Malte and Tran, Duong and Adachi, Masato and Sperl, Matthias},
    title = {Magnetically excited granular matter in low gravity},
    journal = {Review of Scientific Instruments},
    volume = {90},
    number = {5},
    pages = {054501},
    year = {2019},
    month = {05},
    issn = {0034-6748},
    doi = {10.1063/1.5085319},
    url = {https://doi.org/10.1063/1.5085319},
}

@article{shao:20,
	title = {Machine learning holography for {3D} particle field imaging},
	volume = {28},
	doi = {10.1364/OE.379480},
	journal = {Optics Express},
	author = {Shao, Siyao and Mallery, Kevin and Kumar, S. Santosh and Hong, Jiarong},
	year = {2020},
	pages = {2987--2999},
}

@article{niemann:25,
	title = {{ParticleTracking}: {A} {GUI} and library for particle tracking on stereo camera images},
	volume = {10},
	shorttitle = {{ParticleTracking}},
	doi = {10.21105/joss.05986},
	journal = {Journal of Open Source Software},
	author = {Niemann, Adrian and Puzyrev, Dmitry and Stannarius, Ralf},
	year = {2025},
	pages = {5986},
}

@article{puzyrev:24,
	title = {Cooling of a granular gas mixture in microgravity},
	volume = {10},
	doi = {10.1038/s41526-024-00369-5},
	journal = {npj Microgravity},
	author = {Puzyrev, Dmitry and Trittel, Torsten and Harth, Kirsten and Stannarius, Ralf},
	year = {2024},
	pages = {36},
}

@article{dillavou:24,
	title = {Bellybutton: accessible and customizable deep-learning image segmentation},
	volume = {14},
	copyright = {2024 The Author(s)},
	shorttitle = {Bellybutton},
	doi = {10.1038/s41598-024-63906-y},
	journal = {Scientific Reports},
	author = {Dillavou, Sam and Hanlan, Jesse M. and Chieco, Anthony T. and Xiao, Hongyi and Fulco, Sage and Turner, Kevin T. and Durian, Douglas J.},
	year = {2024},
	pages = {14281},
}

@article{Harth2018,
  title = {Free Cooling of a Granular Gas of Rodlike Particles in Microgravity},
  author = {Harth, Kirsten and Trittel, Torsten and Wegner, Sandra and Stannarius, Ralf},
  journal = {Phys. Rev. Lett.},
  volume = {120},
  issue = {21},
  pages = {214301},
  numpages = {6},
  year = {2018},
  month = {May},
  publisher = {American Physical Society},
  doi = {10.1103/PhysRevLett.120.214301},
  url = {https://link.aps.org/doi/10.1103/PhysRevLett.120.214301}
}

@manual{ZarmUserManual2023,
  author = {M. Cornelius},
  title = {ZARM drop tower Bremen User manual},
  date = {July 11, 2023}, 
  year = {2023},
  organization = {ZARM Drop Tower Operation and Service Company},
  pubstate = {July 11, 2023},  
  URI = {https://www.zarm.uni-bremen.de/fileadmin/user_upload/drop_tower/ZARM_BDT_PUG_ver1.3.pdf},  
}

@inproceedings{He2017,
  title={Mask r-cnn},
  author={He, Kaiming and Gkioxari, Georgia and Doll{\'a}r, Piotr and Girshick, Ross},
  booktitle={Proceedings of the IEEE international conference on computer vision},
  pages={2961--2969},
  year={2017}
}

@misc{lin:18,
	title = {Focal {Loss} for {Dense} {Object} {Detection}},
	doi = {10.48550/arXiv.1708.02002},
	publisher = {arXiv},
	author = {Lin, Tsung-Yi and Goyal, Priya and Girshick, Ross and He, Kaiming and Dollár, Piotr},
	year = {2018},
	note = {arXiv:1708.02002 [cs]},
}

@Article{ImageJ2,
author={Rueden, Curtis T.
and Schindelin, Johannes
and Hiner, Mark C.
and DeZonia, Barry E.
and Walter, Alison E.
and Arena, Ellen T.
and Eliceiri, Kevin W.},
title={ImageJ2: ImageJ for the next generation of scientific image data},
journal={BMC Bioinformatics},
year={2017},
volume={18},
pages={529},
doi={10.1186/s12859-017-1934-z},
}

@misc{DigitalSreeni2020,
  author       = {Bhattiprolu, S.},
  title        = {Defining U-net in Python using Keras},
  howpublished = {\url{https://github.com/bnsreenu/python_for_microscopists/blob/master/074-Defining U-net in Python using Keras.py}},
  year         = {2020},
  note         = {Last accessed: Jan 31, 2020}
}

@ARTICLE{watershed_Beucher1979,
  author={Beucher S, Lantuejoul C.},
  journal={International Workshop on Image Processing: Real-time Edge and Motion Detection/Estimation}, 
  title={Use of watersheds in contour detection}, 
  year={1979},
  pages={12-21},
}

@article{liang:23,
	title = {Particle identification in particle tracking velocimetry using two-stage neural networks},
	volume = {19},
	copyright = {http://creativecommons.org/licenses/by/3.0/},
	doi = {10.3934/jimo.2022175},
	journal = {Journal of Industrial and Management Optimization},
	author = {Liang, Jiaming and Liu, Xiaoqi and Chen, Tehuan and Pan, Changchun and Xu, Chao},
	year = {2023},
	pages = {5331--5352},
}

@article{stringer:21,
	title = {Cellpose: a generalist algorithm for cellular segmentation},
	volume = {18},
	copyright = {2020 The Author(s), under exclusive licence to Springer Nature America, Inc.},
	shorttitle = {Cellpose},
	doi = {10.1038/s41592-020-01018-x},
	journal = {Nature Methods},
	author = {Stringer, Carsen and Wang, Tim and Michaelos, Michalis and Pachitariu, Marius},
	year = {2021},
	pages = {100--106},
}

@inproceedings{weigert2022,
  author    = {Martin Weigert and Uwe Schmidt},
  title     = {Nuclei Instance Segmentation and Classification in Histopathology Images with Stardist},
  booktitle = {The IEEE International Symposium on Biomedical Imaging Challenges (ISBIC)},
  year      = {2022},
  doi       = {10.1109/ISBIC56247.2022.9854534}
}

@article{falcon:13,
	title = {Equation of state of a granular gas homogeneously driven by particle rotations},
	volume = {103},
	doi = {10.1209/0295-5075/103/64004},
	journal = {EPL (Europhysics Letters)},
	author = {Falcon, E. and Bacri, J.-C. and Laroche, C.},
	year = {2013},
	pages = {64004},
}

@article{hinton:06,
	title = {Reducing the {Dimensionality} of {Data} with {Neural} {Networks}},
	volume = {313},
	doi = {10.1126/science.1127647},
	journal = {Science},
	author = {Hinton, G. E. and Salakhutdinov, R. R.},
	year = {2006},
	pages = {504--507},
}

@article{cheng:24,
	title = {Unraveling the role of gravity in shaping intruder dynamics within vibrated granular media},
	volume = {7},
	copyright = {2024 The Author(s)},
	doi = {10.1038/s42005-024-01927-9},
	journal = {Communications Physics},
	author = {Cheng, Ke and Hou, Meiying and Sun, Wei and Qiao, Zhihong and Li, Xiang and Lai, Chufan and Yuan, Jinchao and Li, Tuo and Ye, Fangfu and Chen, Ke and Yang, Mingcheng},
	year = {2024},
	pages = {425},
}

@InProceedings{Sudre2017,
author="Sudre, Carole H.
and Li, Wenqi
and Vercauteren, Tom
and Ourselin, Sebastien
and Jorge Cardoso, M.",
editor="Cardoso, M. Jorge
and Arbel, Tal
and Carneiro, Gustavo
and Syeda-Mahmood, Tanveer
and Tavares, Jo{\~a}o Manuel R.S.
and Moradi, Mehdi
and Bradley, Andrew
and Greenspan, Hayit
and Papa, Jo{\~a}o Paulo
and Madabhushi, Anant
and Nascimento, Jacinto C.
and Cardoso, Jaime S.
and Belagiannis, Vasileios
and Lu, Zhi",
title="Generalised Dice Overlap as a Deep Learning Loss Function for Highly Unbalanced Segmentations",
booktitle="Deep Learning in Medical Image Analysis and Multimodal Learning for Clinical Decision Support ",
year="2017",
publisher="Springer International Publishing",
address="Cham",
pages="240--248",
isbn="978-3-319-67558-9"
}

@misc{LabelStudio,
  title={{Label Studio}: Data labeling software},
  url={https://github.com/HumanSignal/label-studio},
  author={
    Maxim Tkachenko and
    Mikhail Malyuk and
    Andrey Holmanyuk and
    Nikolai Liubimov},
  year={2020-2025},
}

\end{document}